\documentclass[a4paper]{article}
\usepackage{listings}
\usepackage{xcolor}
\usepackage{multirow}
\usepackage{jheppub}  
\usepackage{dcolumn}
\usepackage[english]{babel}
\usepackage[utf8x]{inputenc}
\usepackage[T1]{fontenc}
\usepackage{palatino}
\usepackage{amsmath}
\usepackage{afterpage}
\usepackage{graphicx, subcaption}
\usepackage[colorinlistoftodos]{todonotes}
\usepackage[colorlinks=true]{hyperref}
\usepackage{makeidx}
\newcommand{\be}{\begin{equation}}  
\newcommand{\ee}{\end{equation}}  
\newcommand{\bea}{\begin{eqnarray}}  
\newcommand{\eea}{\end{eqnarray}}  
\makeindex
\begin{document}

\vspace*{1.2cm}

\thispagestyle{empty}
\begin{center}
{\LARGE \bf Semiexclusive dilepton production in proton-proton collisions with one forward proton measurement at the LHC}

\par\vspace*{7mm}\par

{

\bigskip

\large \bf Barbara Linek}

\bigskip

{\large \bf   E-Mail: barbarali@dokt.ur.edu.pl}

\bigskip

{College of Natural Sciences, Institute of Physics,
University of Rzesz\'ow, ul. Pigonia 1, PL-35-959 Rzesz\'ow, Poland}

\bigskip

{\it Presented at the Low-$x$ Workshop, Elba Island, Italy, September 27--October 1 2021}

\vspace*{15mm}

\end{center}
\vspace*{1mm}

\begin{abstract}
  aWe discuss the mechanisms of photon-photon fusion triggering the dilepton production in proton-proton collisions with rapidity gap in the main detector 
and one forward proton in the forward proton detectors. 
This correspond to the LHC measurements made by ATLAS+AFP and CMS+PPS. 
Transverse momenta of the intermediate photons and photon fluxes expressed by the form factors and structure functions are included. Moreover the influence of proton measurement and cuts on the $\xi_{1/2}$, $M_{ll}$, $Y_{ll}$, $p_{t,ll}$ distributions, cross section and gap survival factor are considered for both double-elastic and single-dissociative processes. \\
The analyzes used the SuperChic generator to calculate the soft gap survival factor and to compare the results obtained with its use to the results obtained on the owned codes available.
It is shown that the gap survival factor for the single dissociative mechanism is related to the emission of a (mini)jet into the main detector depending on the type of contribution. The dependence between this value and the invariant mass as well as the transverse momentum of the leptons pair and its rapidity was also found.

\end{abstract}
 
 \section{Introduction}
 The photon-photon fusion is one of the dilepton production mechanism in proton-proton collisoins, that was measured recently by the CMS \cite{CMS} and ATLAS \cite{ATLAS} collaborations for the cases with one proton measurement in forward direction. A description of the codes created by our group that enables the analysis of such processes is included in \cite{SFPSS2015, LSS2016, LSS2018}. This is the basis of our formalism to include the gap survival factor associated with the emission of (mini) jets for the production of $W^+ W^-$ \cite{FLSS2019}, $t \bar t$ \cite{LFSS2019} and $\mu^+ \mu^-$ \cite{SLL2021}. However, it is necessary to apply the kinematic condition to the $\xi$-variables (longitudinal momentum fraction loss) in order to compare the theoretical data with the experimental results \cite{CMS,ATLAS}.
 The results of the research from the last mentioned work are presented in this article. 
\begin{figure}
\includegraphics[width=5.1cm]{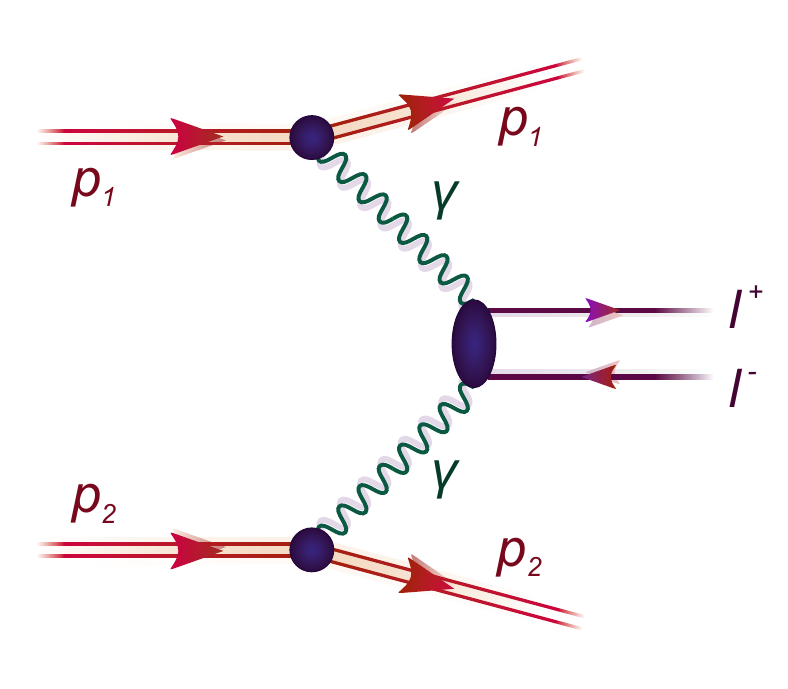}
\includegraphics[width=5.1cm]{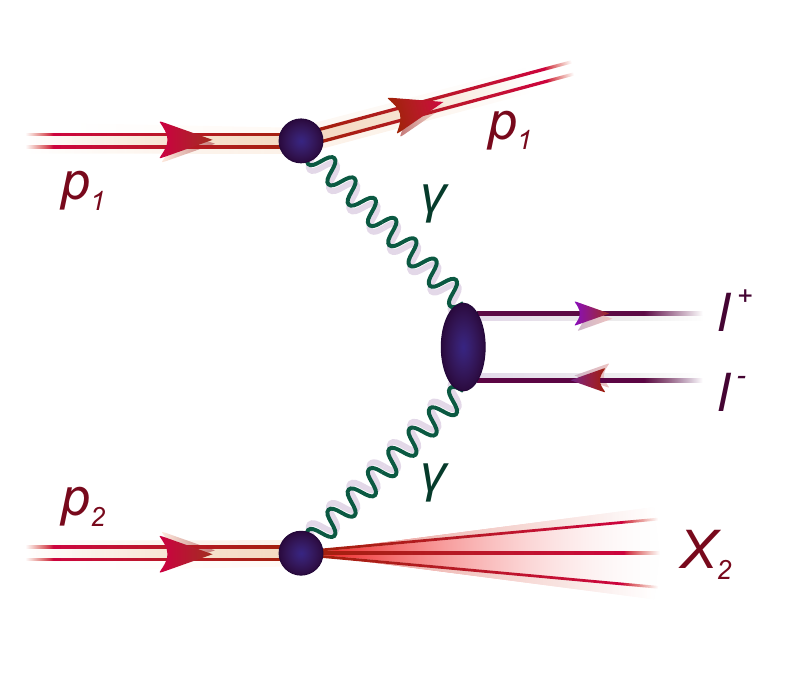}
\includegraphics[width=5.1cm]{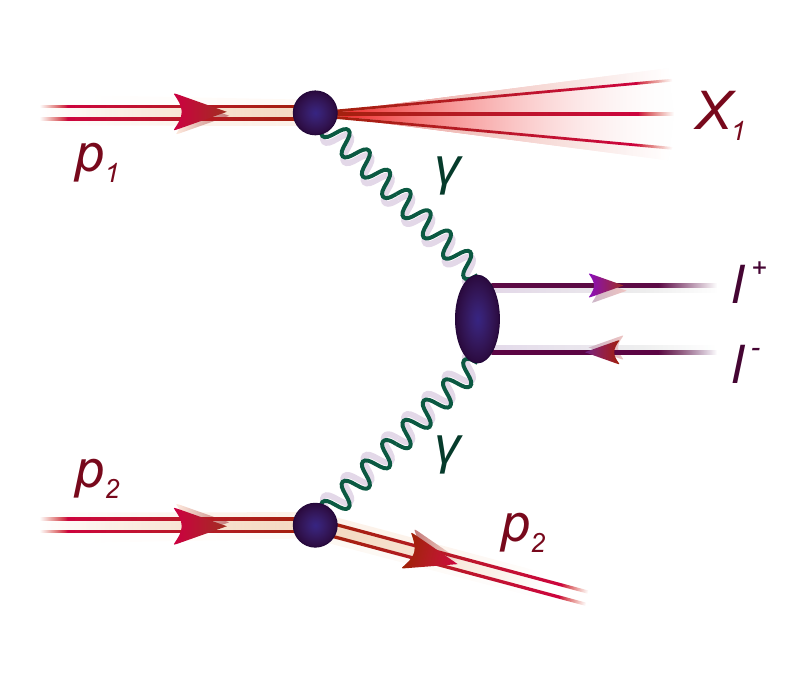}
\caption{Three different categories of $\gamma \gamma$ fusion mechanisms of dilepton production in proton-proton collisions.}
\label{fig:diagrams}
\end{figure}

This analysis is related to the three types of photon initiated events shown in Fig.\ref{fig:diagrams}. There are double elastic and two types of single dissociation processes. The analysis of the mentioned processes is performed with the use of $k_{T}$-factorization approach in which the cross-section for $l^+l^-$ is described as: 
\begin{eqnarray}
\frac{d \sigma^{(i,j)}}{dy_1 dy_2 d^2 {\bf p_1} d^2 {\bf p_2}} &=&  \int  {\frac{d^2 {\bf q_1}}{\pi {\bf q_1^2}} \frac{d^2 {\bf q_2}}{ \pi {\bf q_2^2}} } 
{\cal{F}}^{(i)}_{\gamma^*/A}(x_1,{\bf q_1}) \, {\cal{F}}^{(j)}_{\gamma^*/B}(x_2,{\bf q_2}) 
\frac{d \sigma^*(p_1,p_2;{\bf q_1},{\bf q_2})}{ dy_1 dy_2 d^2{\bf p_1} d^2{\bf p_2}} \, , \nonumber \\ 
\label{eq:kt-fact}
\end{eqnarray}
where the indices $i,j \in \{\rm{el}, \rm{in} \}$ denote elastic or 
inelastic final states.
Here the photon flux for inelastic case is integrated over the mass
of the remnant.

The longitudinal momentum fractions of photons are obtained from 
the rapidities and transverse momenta of final state $l^+l^-$ as:
\begin{eqnarray}
x_1 &=& \sqrt{ {{\bf p_1^2} + m_l^{2} \over s}} e^{+y_1} +  
        \sqrt{ {{\bf p_2^2} + m_l^{2} \over s}} e^{+y_2} 
\; , \nonumber \\
x_2 &=& \sqrt{ {{\bf p_1^2} + m_l^{2} \over s}} e^{-y_1} 
     +  \sqrt{ {{\bf p_2^2} + m_l^{2} \over s}} e^{-y_2} \, .
\end{eqnarray}
The integrated fluxes for elastic and inelastic processes are expressed 
for the elastic case by the proton electromagnetic form factor, while the inelastic flux is expressed by the proton structure function $F_{2}(x_{Bj},Q^{2})$ and $F_{L}(x_{Bj},Q^{2})$ \cite{LSS2016,LSS2018}:

\begin{eqnarray}
\cal F_{\gamma^{*} \leftarrow A}^{\rm{el}}(\rm{z},{\bf q}) &=& \frac{\alpha_{\rm{em}}}{\pi} \left[(1-z) \left(\frac{{\bf q^{2}}}{{\bf q^{2}}+z\left(M_{x}^{2}-m_{A}^{2}\right)+z^{2}m_{A}^{2}} \right)^{2} \frac{4m_{p}^{2}G_{E}^{2}(Q^{2})+Q^{2} G_{M}^{2}(Q^{2})}{4m_{p}^{2}+Q^{2}} \right] 
\;, \nonumber \\
\cal F_{\gamma^{*} \leftarrow A}^{\rm{in}}(\rm{z},{\bf q}) &=& \frac{\alpha_{\rm{em}}}{\pi} \left[(1-z) \left(\frac{{\bf q^{2}}}{{\bf q^{2}}+z\left(M_{x}^{2}-m_{A}^{2}\right)+z^{2}m_{A}^{2}} \right)^{2} \frac{F_{2}(x_{Bj},Q^{2})}{Q^{2}+M_{x}^{2}-m_{p}^{2}} \right] \;.
\end{eqnarray}

Then the four-momenta of intermediate photons are written as:
\begin{eqnarray}
q_1 &\approx& \left( x_1 \frac{\sqrt{s}}{2}, \vec{q}_{1t}, x_1 \frac{\sqrt{s}}{2} \right)
\; , \nonumber \\
q_2 &\approx& \left( x_2 \frac{\sqrt{s}}{2}, \vec{q}_{2t}, -x_2 \frac{\sqrt{s}}{2} \right)
\, .
\end{eqnarray}

\begin{figure}
\begin{center}
\includegraphics[width=6cm]{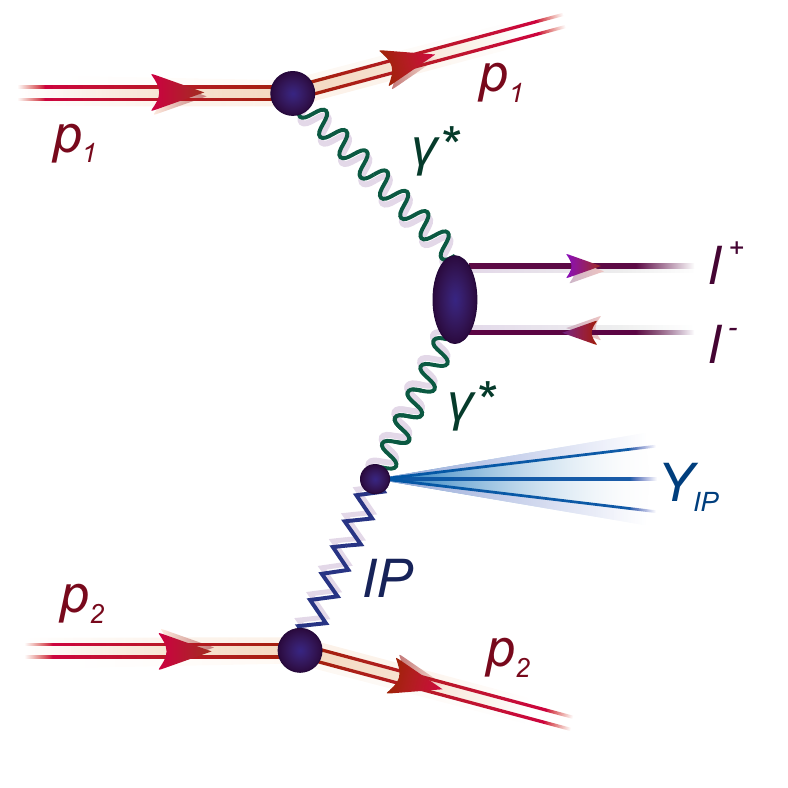}
\includegraphics[width=6cm]{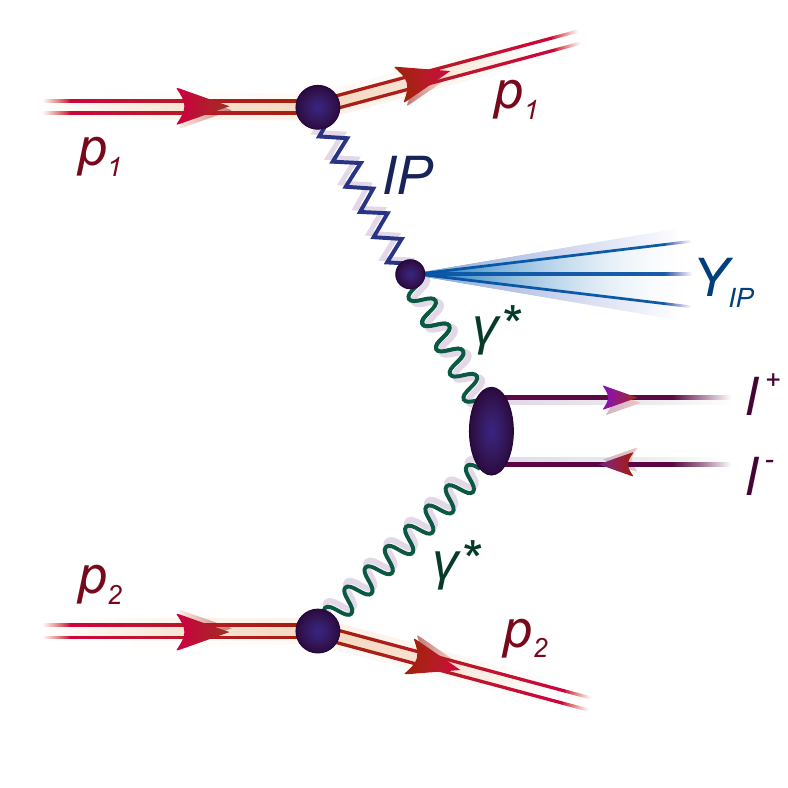}
\caption{Diffractive mechanisms of dilepton production
in proton-proton collisions.}
\label{fig:diffractive_diagrams}
\end{center}
\end{figure}

There is non-zero probability of a proton emission from the remnant system. The analysis of this requires modeling of remnant fragmentation. $\xi$ for this type of processes is more than 0.1 therefore, the measurement in the Roman pots of the ATLAS or CMS experiments is not possible. This codition can be only met by the diffractive mechanism shown in the Fig.\ref{fig:diffractive_diagrams}. 

Comparing the theoretical data with the experimental results requires the imposition of the $\xi$-variables, that is calculated by ATLAS analysis  as:
\begin{equation}
\xi_1 = \xi_{ll}^+  \; , \;  \xi_2 = \xi_{ll}^- \; .
\end{equation}
The longitudinal momentum fractions of the photons were calculated
by them as:
\begin{eqnarray}
\xi_{ll}^+ &=& \left( M_{ll}/\sqrt{s} \right) \exp(+Y_{ll}) \; , \nonumber \\
\xi_{ll}^- &=& \left( M_{ll}/\sqrt{s} \right) \exp(-Y_{ll}) \; .
\end{eqnarray}
The above formula was also used in the discussed analyzes however, only lepton variables are entered in it.

\section{Results}
In the calculations described below were shall take typical cuts on dileptons:
-2.5 $< y_1, y_2 <$ 2.5, $p_{1t}, p_{2t} >$ 15 GeV and extra cuts on $\xi_{ll}^+$ or $\xi_{ll}^-$.
The results from SuperChic \cite{HTKR2020} and from our codes are similar, therefore they are not duplicated. All results are available in \cite {SLL2021}.

\begin{figure}
\includegraphics[width=7cm]{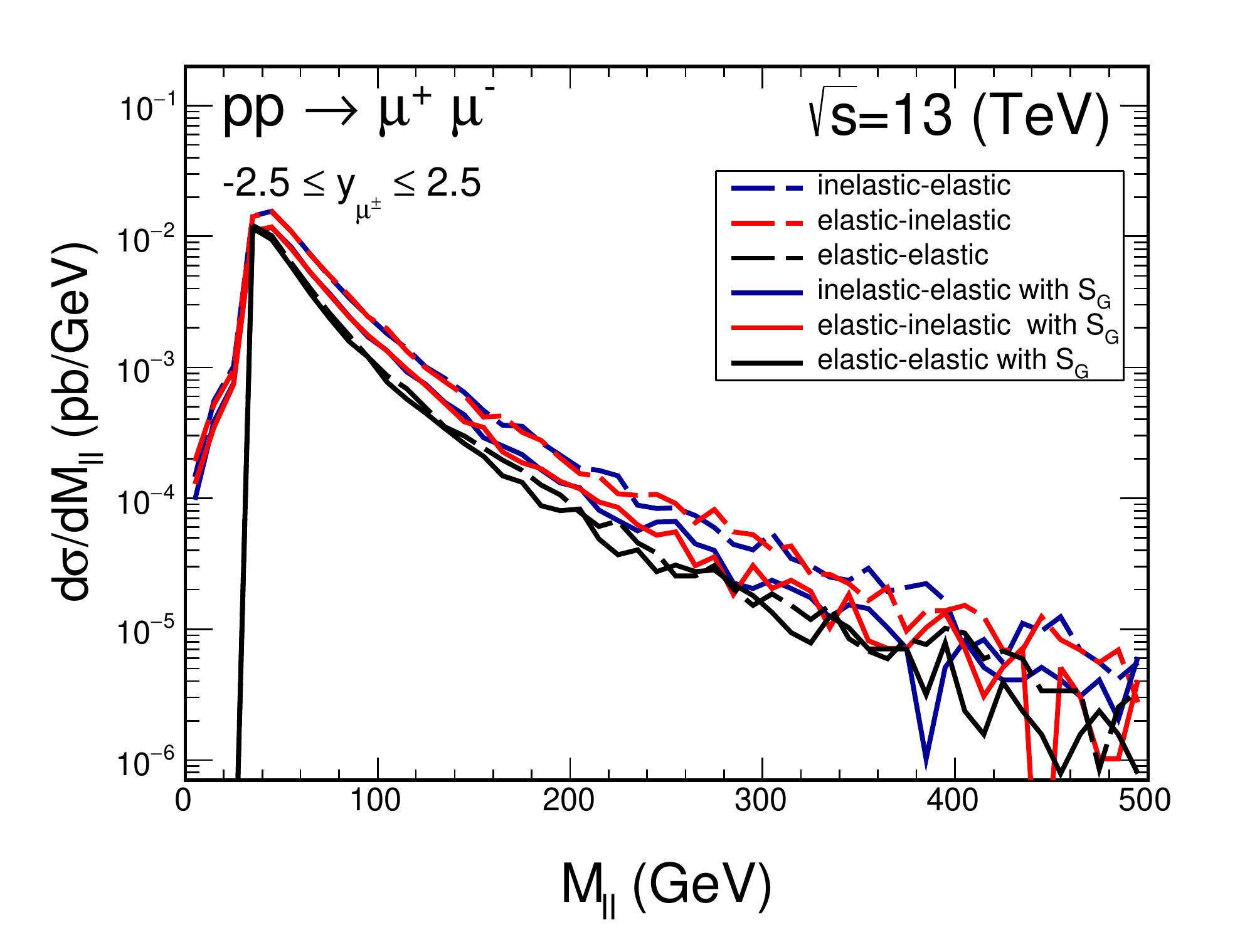}
\includegraphics[width=7cm]{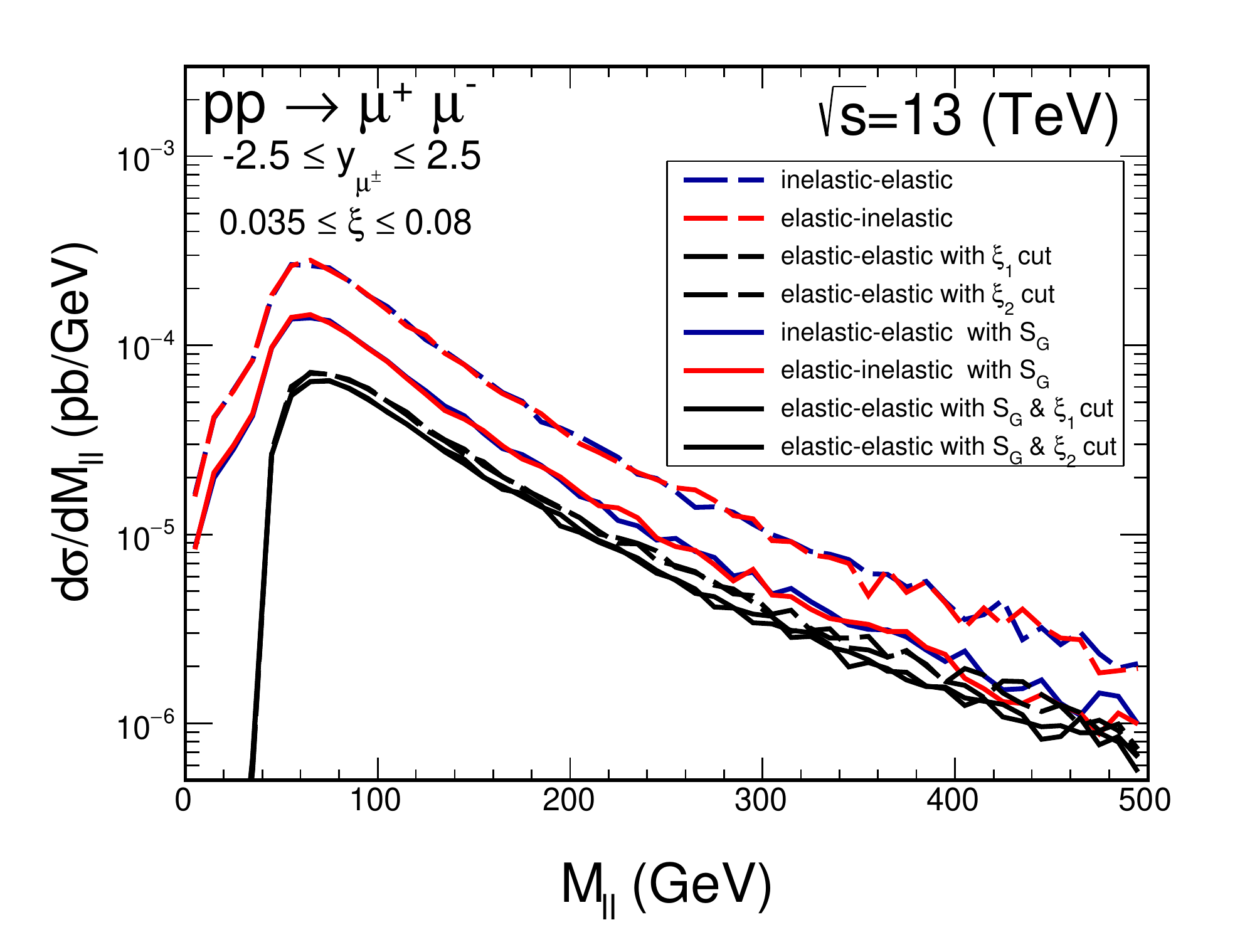}
\caption{Distribution in dimuon invariant mass for
the different contributions considered.
We consider the case without $\xi$ cuts (left panel) and
with $\xi$ cuts (right panel).
}
\label{fig:dsig_dMll_SUPERCHIC}
\end{figure}

In Fig.\ref{fig:dsig_dMll_SUPERCHIC} the distribution 
in dimuon invariant mass for the case without $\xi$ cuts (left panel) and with $\xi$ cuts (right panel) are presented. The elastic-elastic (dashed line)
and elastic-inelatic + inelastic-elastic (solid line) are shown separately.
On average larger invariant masses in the case with $\xi$ cuts are observed.

\begin{figure}
\includegraphics[width=7cm]{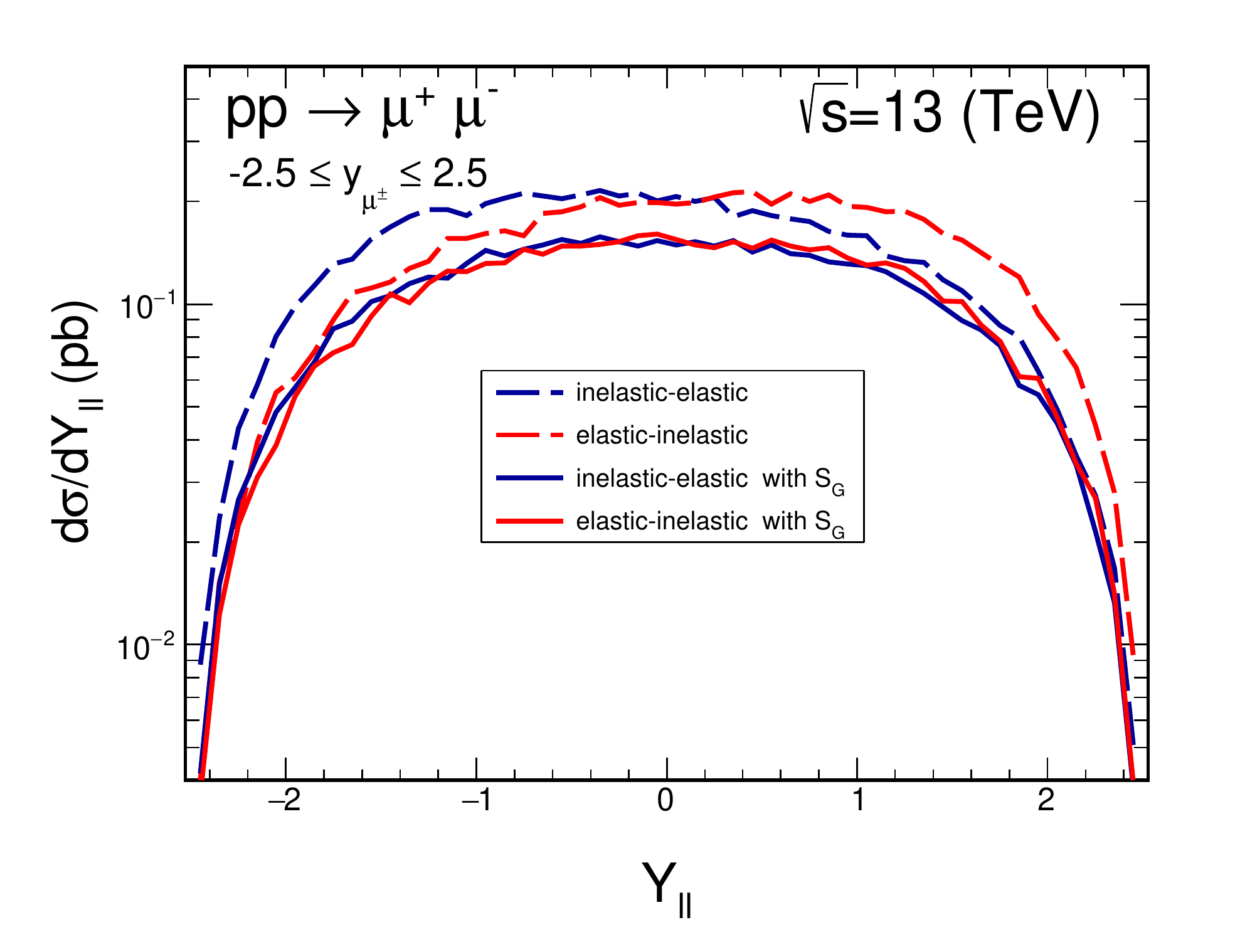}
\includegraphics[width=7cm]{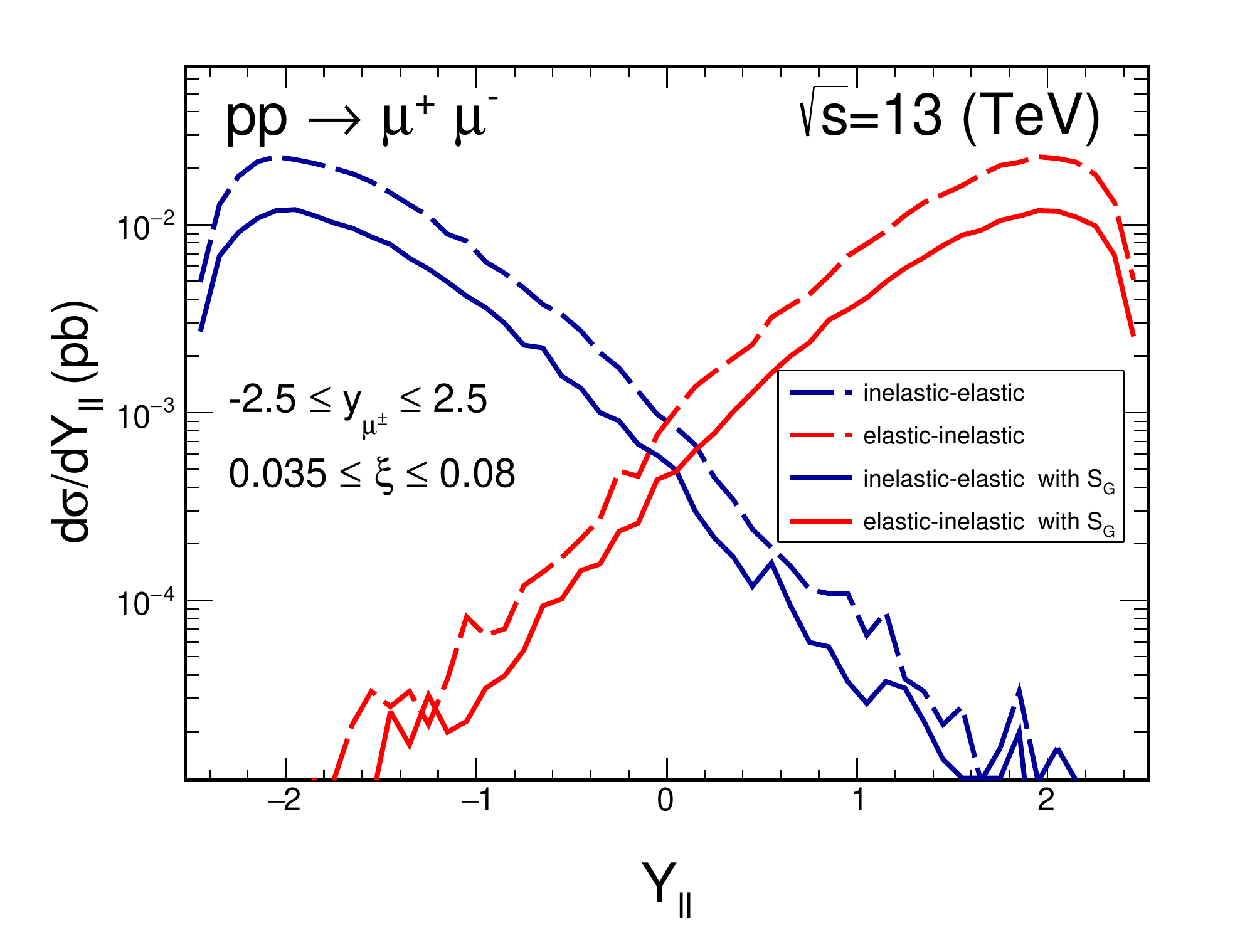}
\caption{Distribution in rapidity of the dimuon pair.
We show the case without $\xi$ cuts (left panel) and
with $\xi$ cuts (right panel).
for the different contributions considered.
}
\label{fig:dsig_dYll_SUPERCHIC}
\end{figure}

The Fig.\ref{fig:dsig_dYll_SUPERCHIC} presents similar
distributions but in $Y_{ll}$.
Without the $\xi$ cut quite different shapes of distributions
in $Y_{ll}$ without and with soft rapidity gap survival factor 
(see the left panel) can be observed.
When the $\xi$-cut is imposed, the distributions with and without
soft rapidity gap survival factor have very similar shapes.
Then, however, the elastic-inelastic and inelastic-elastic
contributions are well separated in $Y_{ll}$. The sum of both
contributions has a characteristic dip at $Y_{ll}$ = 0. 

\begin{figure}
\includegraphics[width=7cm]{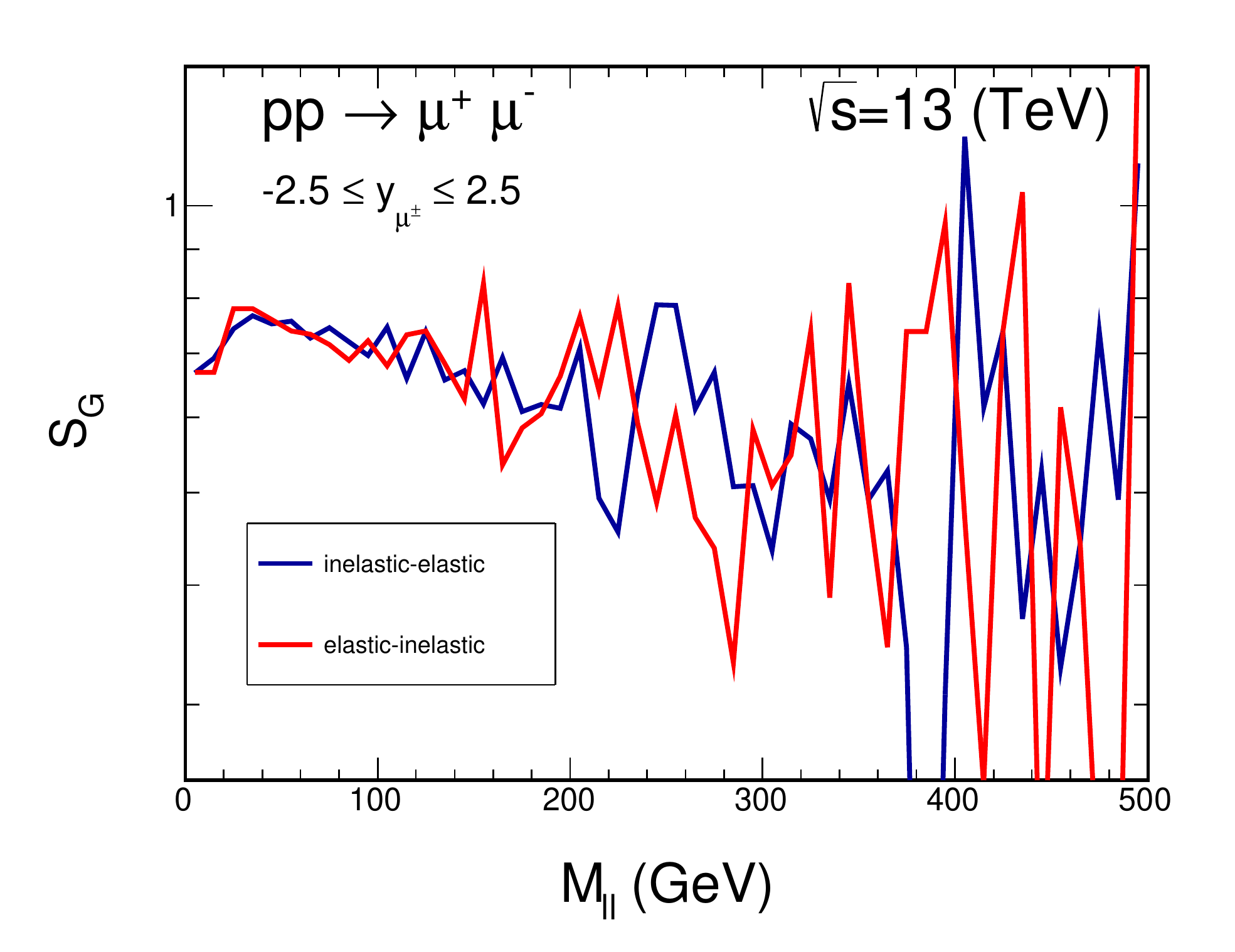}
\includegraphics[width=7cm]{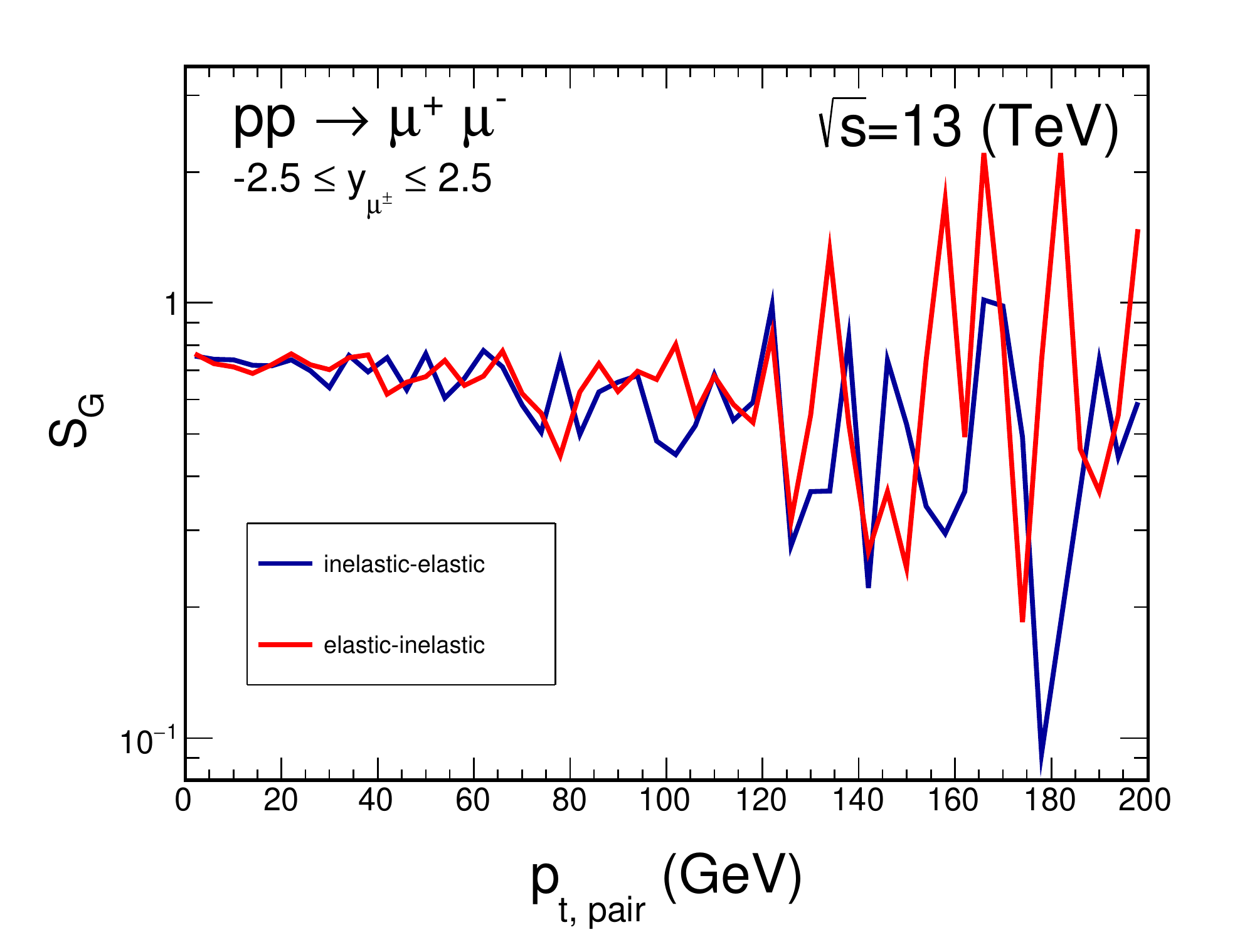}\\
\includegraphics[width=7cm]{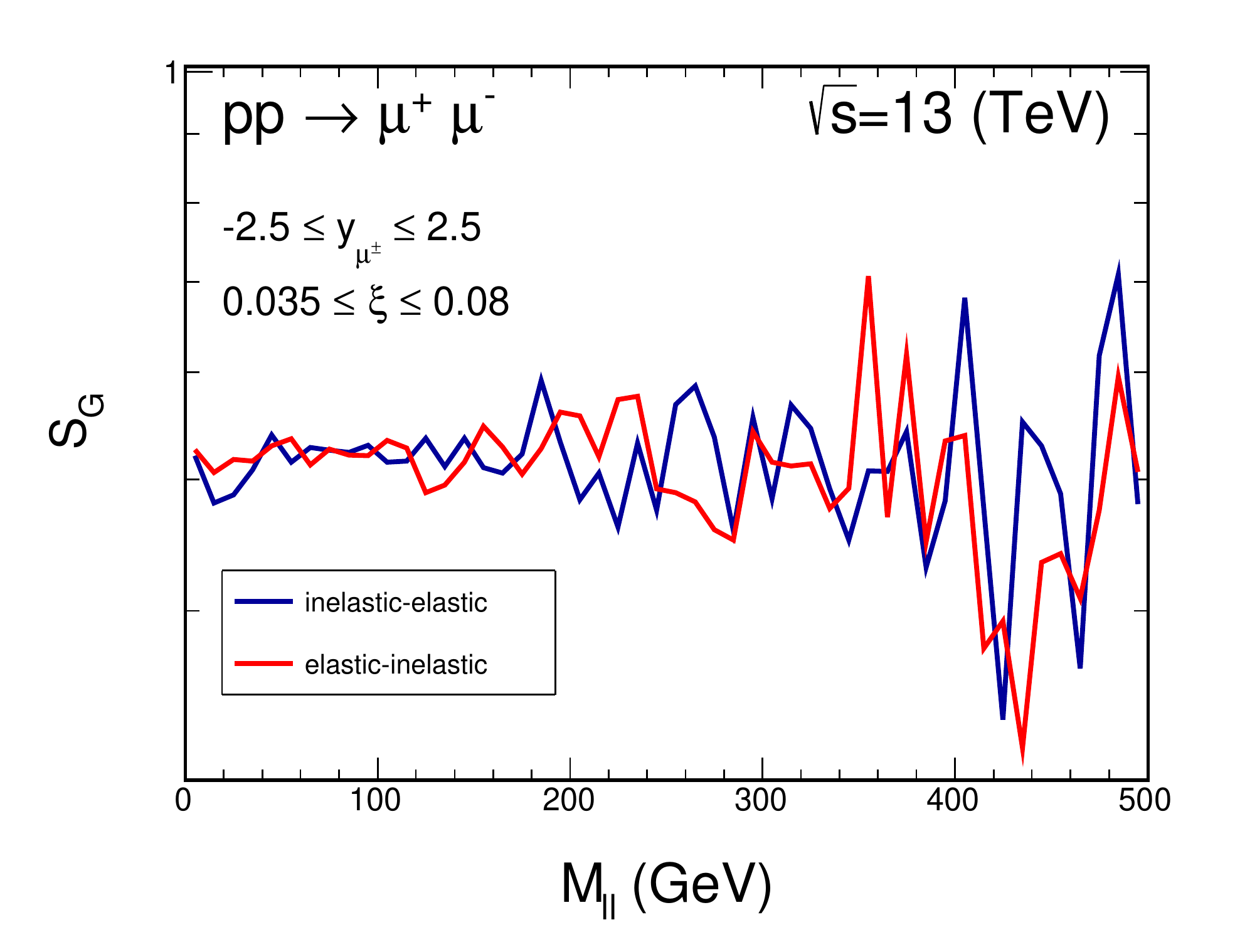}
\includegraphics[width=7cm]{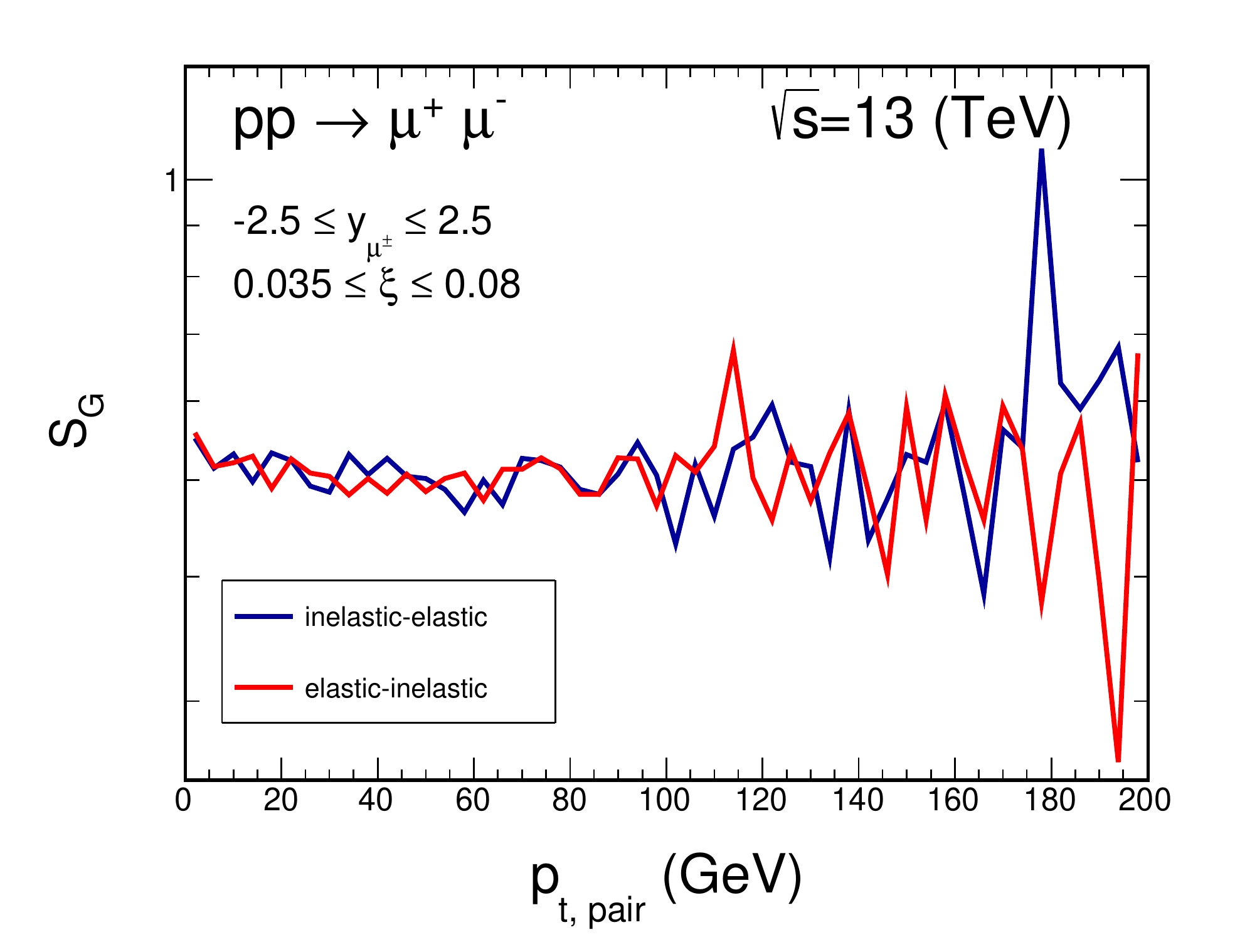}\\
\caption{The soft gap survival factor as a function of
dilepton invariant mass (left panels) and as a function of transverse
momentum of the pair (right panels) for 
single dissociation (solid line) mechanisms.
We show the result without $\xi$ cuts (upper panels) and
with $\xi$ cuts (lower panels).
}
\label{fig:soft_gap_survival_factor}
\end{figure}

The Fig.\ref{fig:soft_gap_survival_factor} shows corresponding
gap survival factor calculated as:
\begin{eqnarray}
S_G(M_{ll}) &=& \frac{d \sigma / d M_{ll}|_{with SR}}
                     {d \sigma / d M_{ll}|_{without SR}}
\; , \\
S_G(p_{t,pair}) &=& \frac{d \sigma / d p_{t,pair}|_{with SR}}
                         {d \sigma / d p_{t,pair}|_{without SR}}
\; .
\label{differential_gap}
\end{eqnarray}
The ratio of the cross section with the soft rapidity gap survival
factor to its counterpart without including the effect was calculated,
the difference is visible in $M_{ll}$ (left panel)
or in $p_{t,pair}$ (right panel) for double elastic (dashed line)
and single dissociation (solid line).
A small dependence on both $M_{ll}$ and on $p_{t,pair}$ can be observed here. 
The gap survival factor for double elastic component
is larger than for single dissociation and the gap survival factor corresponding to the measurement of one proton 
is significantly smaller than that for the inclusive case.
The rather large fluctuations are due to limited statistics (50 000 events).

\begin{figure}
\includegraphics[width=7cm]{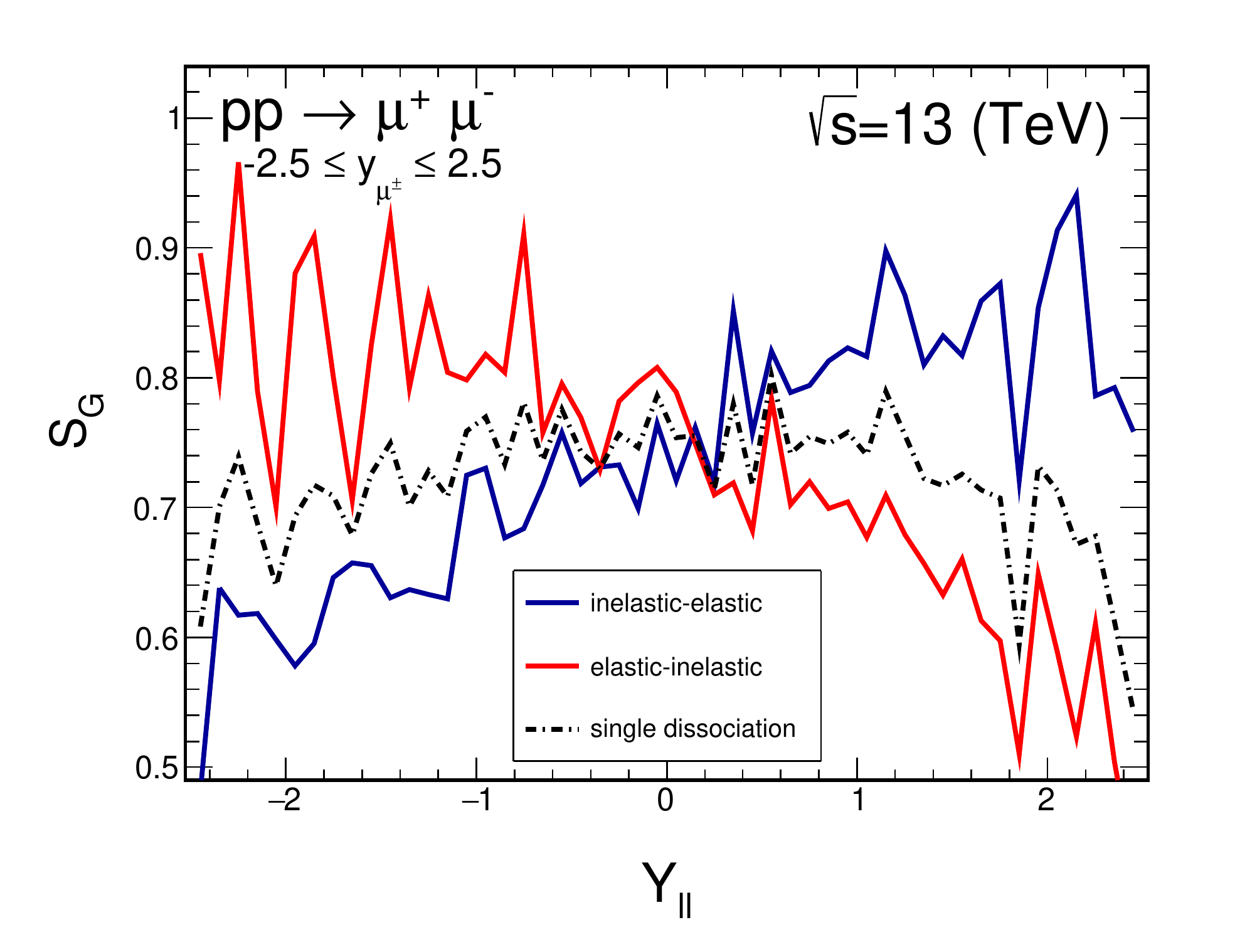}
\includegraphics[width=7cm]{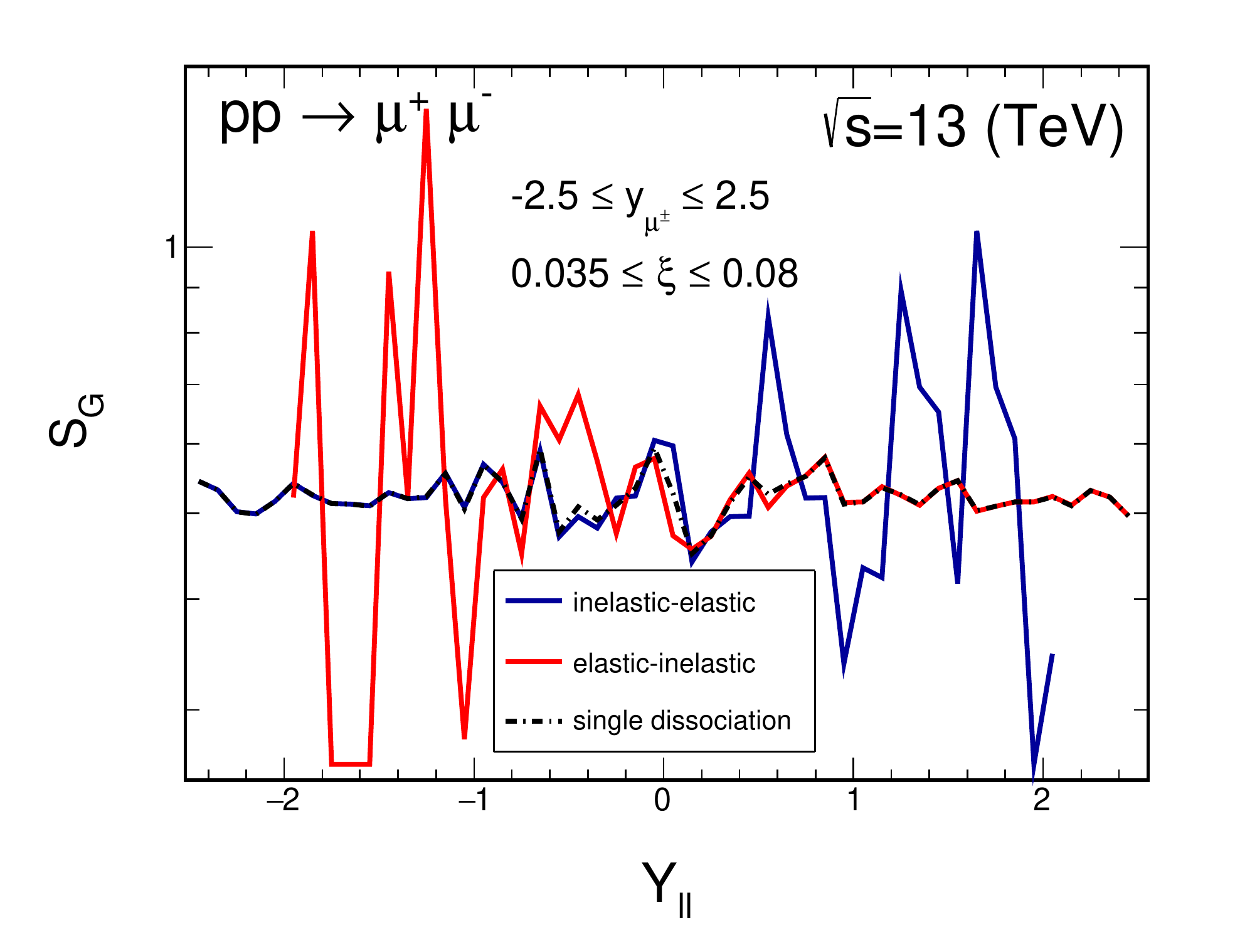}
\caption{The soft gap survival factor as a function of
rapidity of the $\mu^+ \mu^-$ pair for single proton dissociation.
We show the result without $\xi$ cuts (left panel) and
with $\xi$ cuts (right panel). The dash-dotted black line represents
effective gap survival factor for both single-dissociation components
added together.
}
\label{fig:soft_gap_survival_factor_2}
\end{figure}

The Fig.\ref{fig:soft_gap_survival_factor_2} shows in addition 
soft gap survival factor as a function of the rapidity of the dimuon pair.
A strong dependence of the gap survival factor on $Y_{ll}$
separately for elastic-inelastic and inelastic-elastic components can be observed
but only in the case when proton is not measured. This effect may
be very difficult to address experimentally as in this 
(no proton measurement) case one measures the sum of the both (all)
components, where the effect averages and becomes more or less
independent of $Y_{ll}$ (see black dash-dotted curve).
However, it seems interesting to understand the dependence on $Y_{ll}$ 
for individual component from theoretical point of view.\\

\begin{figure}
\includegraphics[width=7cm]{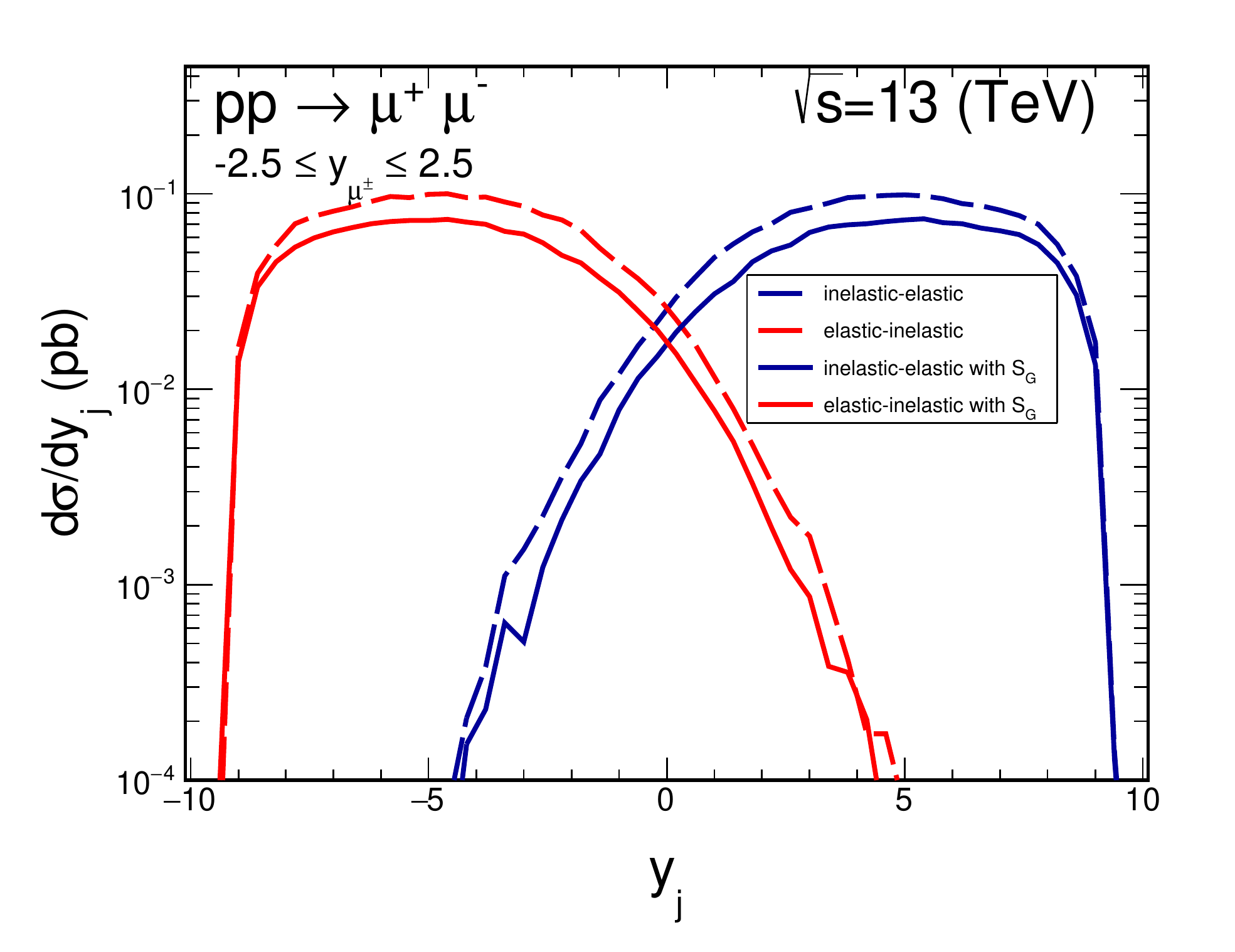}
\includegraphics[width=7cm]{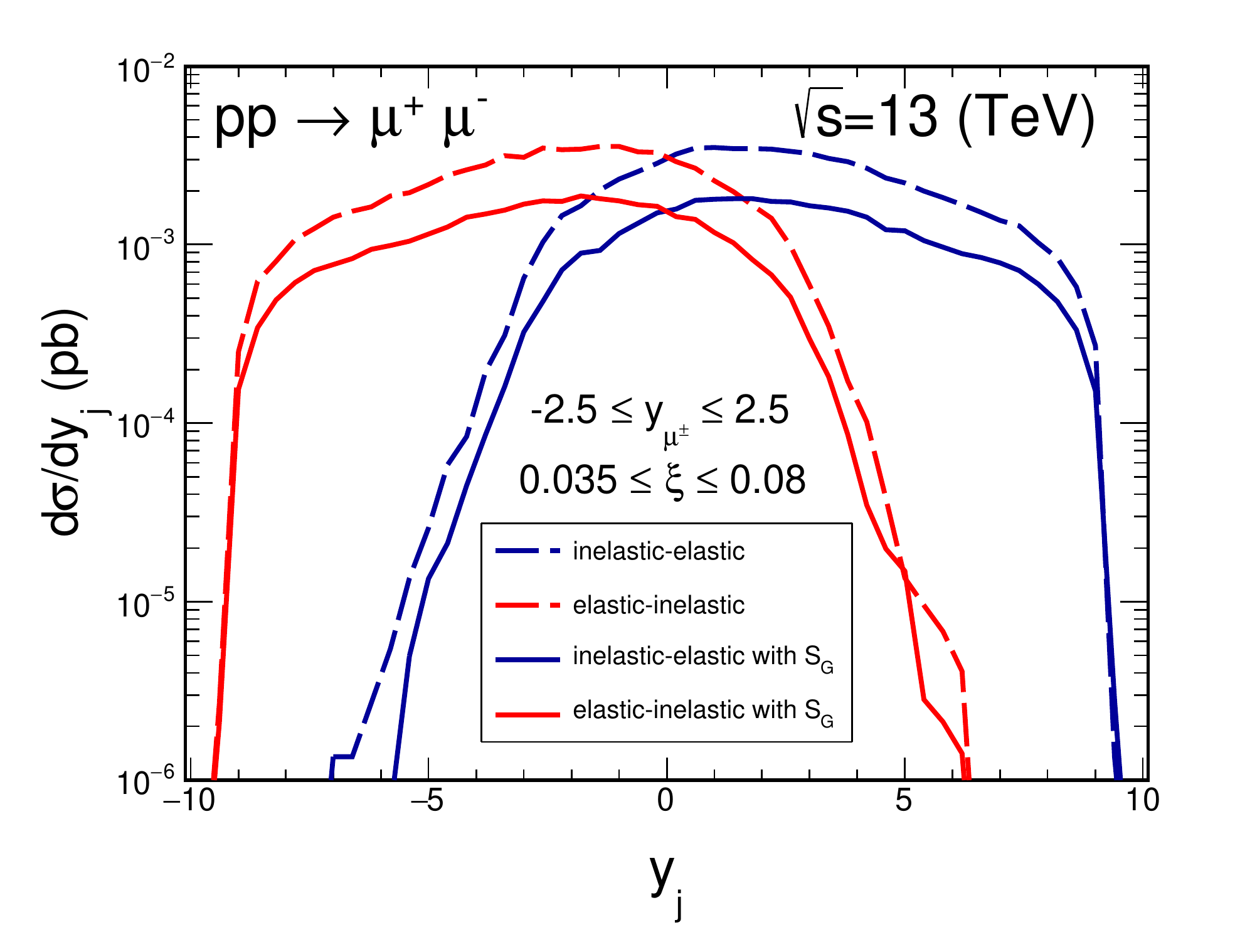}
\caption{Distribution in the (mini)jet rapidity for the inclusive case
with no $\xi$ cut (left panel) and when the cut on $\xi$ is imposed
(right panel) for elastic-inelastic and inelastic-elastic contributions
as obtained from the SuperChic generator.
We show result without (dashed line) and with (solid line) soft
rescattering correction.
}
\label{fig:dsig_dyjet_SUPERCHIC}
\end{figure}

The fact how the proton dissociation further reduces
the gap survival factor due to emission of a (mini)jet that can enter
into the main detector and destroy the rapidity gap is also interesting.
This was discussed e.g. in \cite{HKR2016,LSS2018,LFSS2019}.
In Fig.\ref{fig:dsig_dyjet_SUPERCHIC} the (mini)jet distribution
in rapidity for elastic-inelastic and inelastic-elastic components are shown
 without imposing the $\xi$ cut (left panel)
and when imposing the $\xi$ cut (right panel).
One can observe slightly different shape for both cases.
The corresponding gap survival factor (probability of no jet in the main
detector) is 0.8 and 0.5, respectively.
The probability of no emission around the 
$\gamma \gamma \to \mu^+ \mu^-$ vertex is, however, much more difficult
to calculate and requires inclusion of remnant hadronization
which is model dependent.

In Table I the probability that the (mini)jet is
outside the main detector, i.e.: $y_{jet} <$ -2.5 or $y_{jet} >$ 2.5 is shown.
Imposing cuts on $\xi$ lowers the corresponding (minijet) rapidity 
gap survival factor while imposing extra cut $p_{t,pair} <$ 5 GeV, 
as in the ATLAS experiment, increases it back.

\begin{table}
   \centering
   \caption{Gap survival factor due to minijet emission.
The first block is with only internal SuperChic cut:
-2.5 $< Y_{ll} <$ 2.5, the second block is when the condition
on individual rapidities is imposed extra,
the third block includes in addition the cut on $\xi_1$ or $\xi_2$,
and the final block includes also the condition $p_{t,pair} <$ 5 GeV.
In all cases $p_{1t}, p_{2t} >$ 15 GeV.
In the last panel (*) means 10 000 events only.} 

\begin{tabular}{|c|c|c|}
\hline
contribution & without $S_G$ & with $S_G$ \\
\hline
cut on $Y_{ll}$ only &  & \\
\hline
elastic-inelastic & 0.76304 & 0.78756 \\
inelastic-elastic & 0.76278 & 0.78898 \\ 
\hline
cut on $y_1$ and $y_2$ in addition &  & \\
\hline
elastic-inelastic & 0.77366 & 0.79250 \\  
inelastic-elastic & 0.76926 & 0.78744 \\
\hline
cut on $\xi_1$ or $\xi_2$ in addition &   & \\
\hline
elastic-inelastic & 0.52430 & 0.53976 \\
inelastic-elastic & 0.53118 & 0.53614 \\
\hline
cut on $p_{t,pair}$ in addition &   &   \\
\hline
elastic-inelastic & 0.83144 & 0.84350(*) \\
inelastic-elastic & 0.83462 & 0.84960(*) \\ 
\hline
\end{tabular}
\label{table:jet-survival-factor}
\end{table}

The factor below is included in the case when rapidity
gap condition is imposed experimentally. It is less clear what to do
when the condition of separated $\mu^+$ and $\mu^-$ are imposed
as in the ATLAS experiment \cite{ATLAS}.
In the following it is assume that the particles from (mini)jet, emitted 
from the same vertex as leptons, will always break the conditions, 
provided they are emittted in the same range of rapidities as 
the measured leptons. 
This range is defined by the geometry of the main ATLAS (CMS) detector.

\section{Conclusions}
These proceedings discuss the production of dileptones initiated by a photon-photon fusion with one forward proton measurement. Conditions on $\xi_{1}$ or $\xi_{2}$ for the forward emitted protons were imposed therein. Particulary interesting is the distribution in $M_{ll}$ and in $Y_{ll}$ which has a minimum at $Y_{ll} \approx 0$.Several distributions were discussed in \cite{SLL2021}.

The soft rapidity gap survival factor was caltulated as a function of $M_{ll}$, $p_{t pair}$ and $Y_{ll}$. There are no obvious dependences on the variables have been found for the single dissociation, exept of distribution in $Y_{ll}$, however there is a strong dependence on the proton measurement. 

There was also calculated gap survival factor due to (mini)jet emission by checking whether the (mini)jet enters to the main detector or not. This type of gap survival also strongly depends on the outgoing proton is measurement or not. It is about 0.8 for inclusive case and about 0.5 for the case with proton measurement in forward proton detector.

\end{document}